\documentclass[english]{article}
\pdfoutput=1
\usepackage{geometry}
\usepackage{babel}
\usepackage{xcolor}
\usepackage{graphicx}
\usepackage{amsmath,amsfonts}
\usepackage{subfigmat}
\usepackage{hyperref}
\hypersetup{colorlinks, citecolor=red, linkcolor=blue, urlcolor=red, breaklinks=true}

\begin{document}
\newtheorem{lemma}{Lemma}
\newtheorem{corollary}{Corollary}

 \title{\textbf{An Aerodynamic Analysis of a Robustly Redesigned Modern Aero-Engine Fan} \footnote{Copyright 2016 by Rolls-Royce plc.} }

\author{Pranay Seshadri\footnote{Postdoctoral Fellow, Department of Engineering, Trumpington Street}, Geoffrey T. Parks\footnote{Reader in Engineering, Department of Engineering, Trumpington Street}, Shahrokh Shahpar\footnote{Rolls-Royce Associate Engineering Fellow, CFD Methods, Design Systems Engineering, Moore Lane, AIAA Associate Fellow}}
\date{\textit{University of Cambridge, Cambridge CB2 1PZ, United Kingdom \\ Rolls-Royce plc, Derby DE24 8BJ, United Kingdom}}
\maketitle

\begin{abstract}
This paper documents results from a recent computational study aimed at de-sensitizing fan stage aerodynamics---in a modern, high bypass ratio aero-engine---to the effects of rear-seal leakage flows. These flows are the result of seal erosion between a rotor and stator disk in an engine, and deterioration over the life of an engine. The density-matching technique for optimization under uncertainty was applied to this problem. This involved RANS and adjoint flow solves of a full fan stage carried out at two different leakage conditions. Here a detailed analysis of the fan stage aerodynamics is carried out to determine why exactly the new design is more insensitive to the effects of leakage flows. Specifically, it is shown that this insensitivity is  attributed to three main factors: a slight rearward shift in loading, and thus a reduction in incidence; a reduction in the cross-passage pressure gradient; and a re-acceleration of the flow towards the trailing edge, which prevented any corner separation. 
\end{abstract}

\section*{Nomenclature}
\noindent
\begin{tabular}{@{}lcl@{}}
\textit{$\Gamma$} && circulation\\
\textit{$\gamma$} && universal gas constant\\
\textit{$\theta$} && angle between camberline at the leading edge and $r$\\
\\
\textit{ESS} && engine stator sections \\
\textit{OGV} && outlet guide vane \\
\textit{SFC} && specific fuel consumption \\
\textit{$D_{spike}$} && leading edge spike height\\
\textit{$\dot{m}$} && massflow rate\\
\textit{$P_{0N}$} && stagnation pressure at station $N$\\
\textit{$P_{N}$} && static pressure at station $N$\\
\textit{$r$} && distance from center of circulation to leading edge\\
\textit{$T_{0N}$} && stagnation temperature at station $N$\\
\textit{$T_{N}$} && static temperature at station $N$\\

\end{tabular}

\section{Introduction}
\begin{figure}
  \centerline{\includegraphics[natwidth=962, natheight=492, width=13cm]{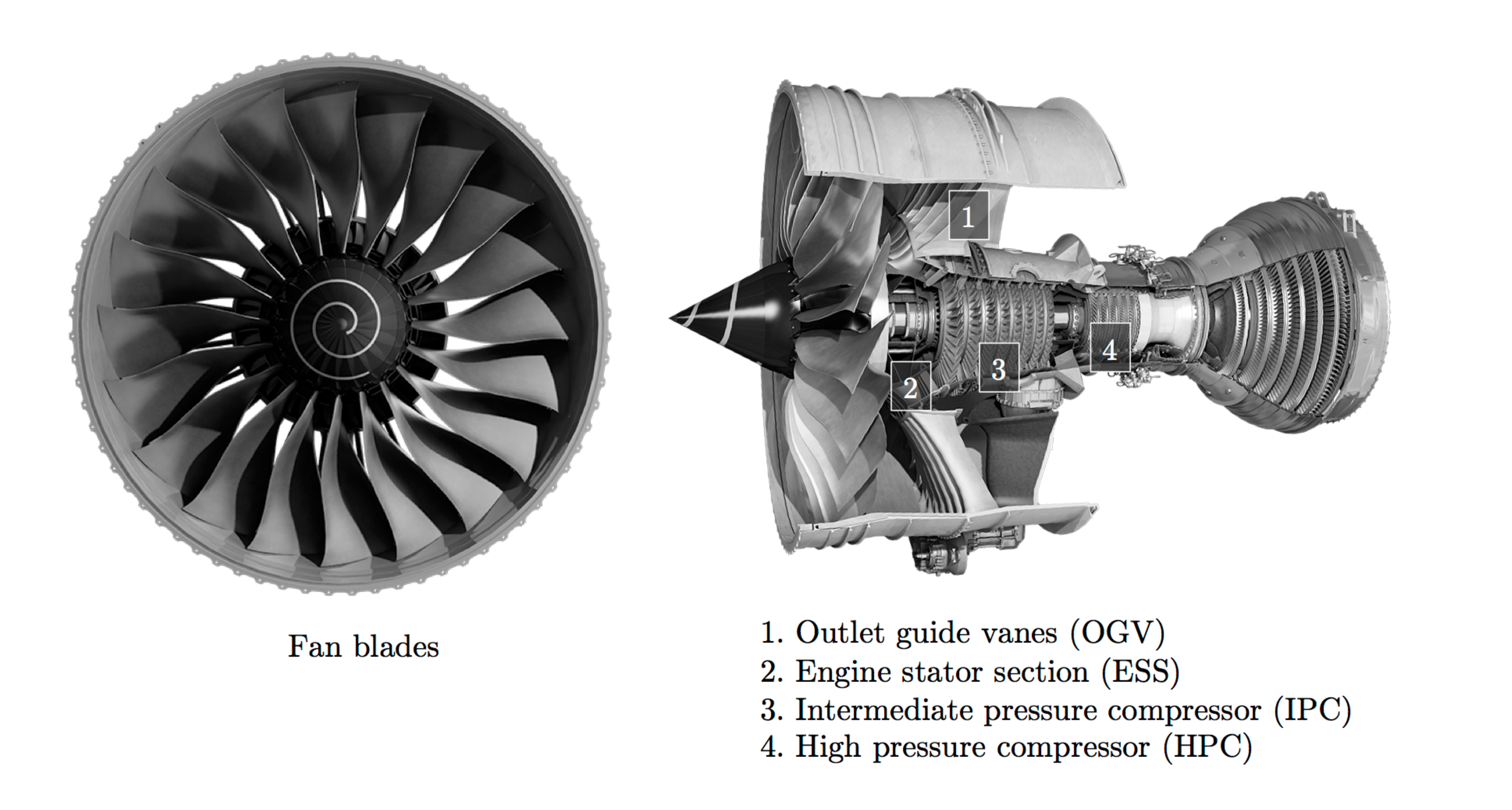}}
  \caption{A Rolls-Royce modern, high bypass ratio, large civil aero-engine. Printed with permission from Rolls-Royce plc.}
\label{fig:engine}
\end{figure}

\begin{figure}
  \centerline{\includegraphics[natwidth=718, natheight=492, width=11cm]{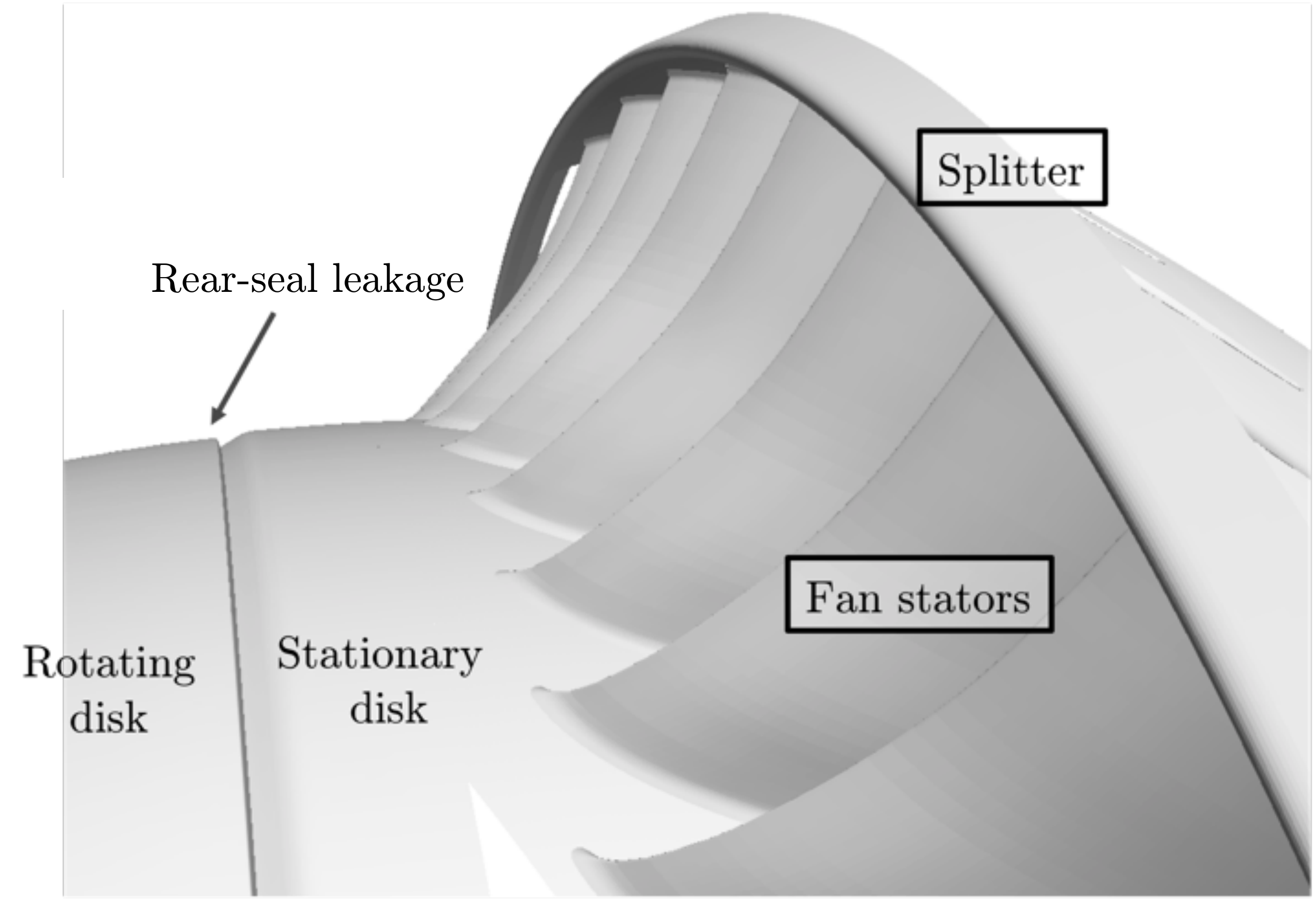}}
  \caption{Origin and location of rear-seal leakage flow}
\label{fig:engine2}
\end{figure}

The fan is the most visible part of a modern aero-engine, providing most of its thrust. The fan is part of the fan sub-system (shown in figure \ref{fig:engine}), which  comprises the fan blades, casing, bypass outlet guide vanes (OGVs) and the engine sector stators (ESS). The fan---also known as the fan rotor blade---is fastened to a root structure, which is attached to the fan disc. This rotating disk is mounted on the fan shaft, which is connected to and driven by the low pressure turbine. Depending on the speed at which this shaft rotates, approximately nine-tenths of the air brought in by the fan, passes through the bypass---to deliver engine thrust---while the remainder goes to the core---for further compression followed by combustion, by the combustor, and expansion, by the turbine \cite{RR}. Between the rotating disk on which the fan blades are cantilevered and the stationary disk that houses the ESS and OGV, there is a small gap. This small gap exists to accommodate rotor disk centrifugal growths and minor axial movements (see figure \ref{fig:engine2}). To prevent air immediately downstream of the fan rotor from leaking out of this gap---thereby reducing the capacity (a function of mass-flow rate) for the core compressor---the gap is pressure sealed using higher pressure air from the core compressor. As a consequence, there is a small rear-seal leakage flow that emanates into the fan stator. This leakage flow---which is a high-pressure (relative to the fan exit flow-field), high-temperature and low-momentum fluid---has been found to act as a catalyst in separating the flow along the suction surface of the fan stator \cite{Zamboni_thesis}. Furthermore, over the life of an engine, this leakage flow may increase and this significantly impacts stator and core compressor aerodynamics. De-sensitizing fan sub-system aerodynamics to the effect of leakage flow deterioration is extremely desirable because the fan sub-system has an extremely favorable efficiency to specific fuel consumption (SFC) ratio, with a 1$\%$ increase in efficiency yielding a $0.7\%$ drop in SFC \cite{Zamboni}. 

There is a dearth of studies in open literature that provide a detailed account of the fluid mechanics of rear-seal leakage flows in fan stages, even though this is an extremely important part of the engine. One of the few studies in this area is the work of \cite{Zamboni} where the authors carry out a parametric study on fan leakage flow parameters and compare high bypass ratio rig tests with computational simulations. They conclude that increasing rear-seal leakage mass-flow rate and stagnation total temperature increase the extent of flow separation observed on the fan stator. They also state that varying the stator endwall curvature can reduce this effect; however, in their work no optimization or design under uncertainty study was carried out. This motivated a prior study \cite{Seshadri_ASME2} by the authors where a new approach to turbomachinery optimization under uncertainty was developed and applied to address the fan rear-seal leakage problem. The approach---titled monotonic density matching---was an outgrowth of previous work in \cite{Seshadri}, which can be interpreted as an adjoint-enhanced method for robust design optimization.

\begin{figure}
  \centerline{\includegraphics[natwidth=663, natheight=445, width=14cm]{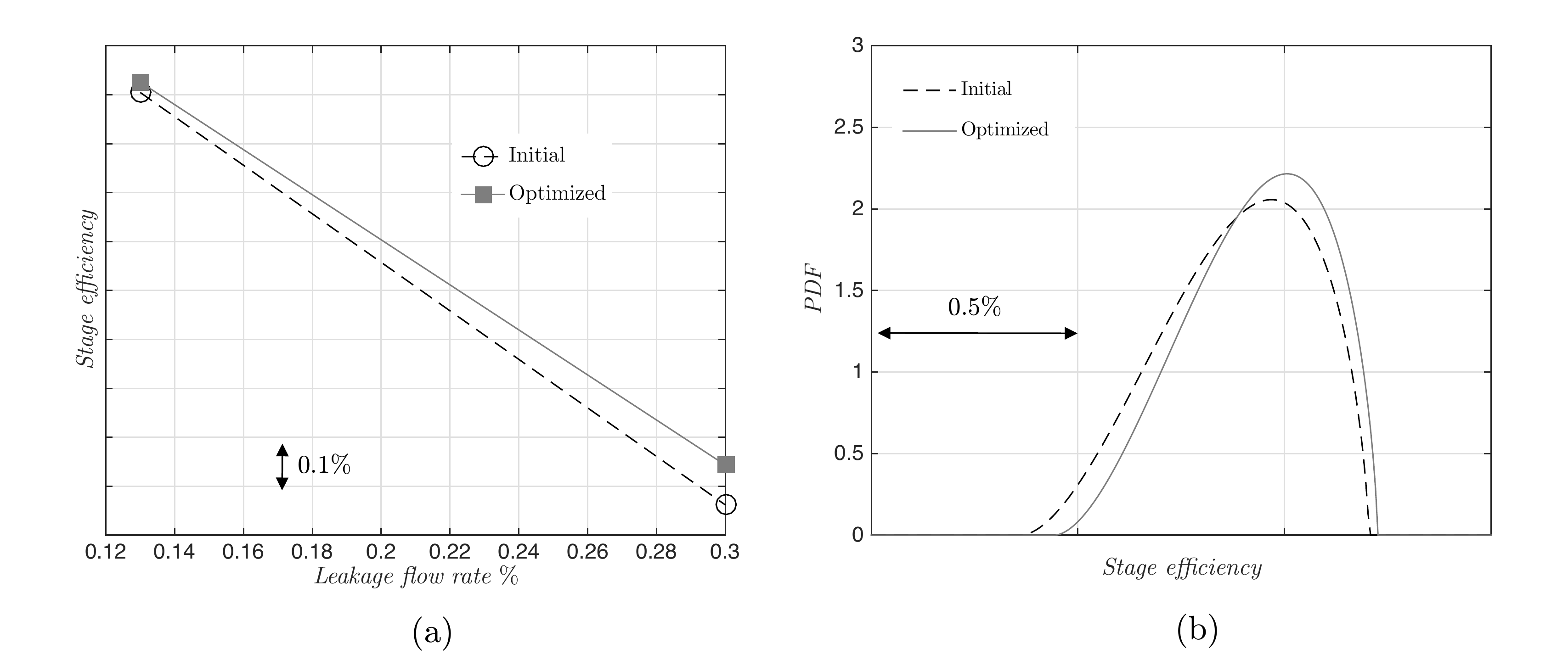}}
  \caption{Fan root stage efficiency with increasing leakage flow: (a) Exchange curves; (b) Corresponding probability density functions, assuming an uncertainty in the rear-seal leakage mass-flow rate. Adapted from \cite{Seshadri_ASME2}}
\label{fig:old_result}
\end{figure}

In \cite{Seshadri_ASME2}, the authors formulate and test their robust design optimization approach on the fan stage by characterizing the rear-seal leakage mass-flow rate as the uncertainty, and the fan root stage efficiency as the quantity-of-interest. The final result of their optimizations is shown in figure \ref{fig:old_result}. Here, (a) plots the variation of fan stage efficiency against increasing rear-seal leakage mass-flow rate, while (b) plots the corresponding probability density functions. Here, the assumed probability distribution in the rear-seal leakage was based on the expected variation of that quantity over the life of an engine. The minimum and maximum values for the rear-seal leakage mass-flow were $0.13\%$ and $0.30\%$ of the stator inlet flow. While the authors obtain a relatively small increase in the efficiency, what is noteworthy is the reduction in the variance. In that paper, the authors simply use the fan stage as a test case, and thus do not elaborate on the underlying aerodynamics. This brings us to the motivation of this paper. The initial design used in \cite{Seshadri_ASME2} is from a modern aero-engine and is very efficient. Thus, obtaining an increase in its mean efficiency---which is a challenge even for today's start-of-the-art optimization algorithms---along with a reduction in its variance, is worth further investigation. Specifically, in this paper, the aim is to isolate the fluid mechanics phenomena responsible for the reduction in variance, and discuss how the `robust design' achieves this. This paper is structured as follows. Section 2 provides a brief literature review of fan and compressor flows, and in section 3, a detailed analysis of the aerodynamics is presented. This is followed by conclusions in section 4.

\section{Literature review}
This literature review is focused on two key concepts: compressor spikes and compressor 3D flows. The former serve as a performance criterion that will be used later in this text when analyzing the robust aerodynamics, while the latter topic is of relevance given the 3D nature of the flow in the fan stage considered.  

\subsection{Compressor spikes}
\begin{figure}
  \centerline{\includegraphics[natwidth=705, natheight=416, width=11cm]{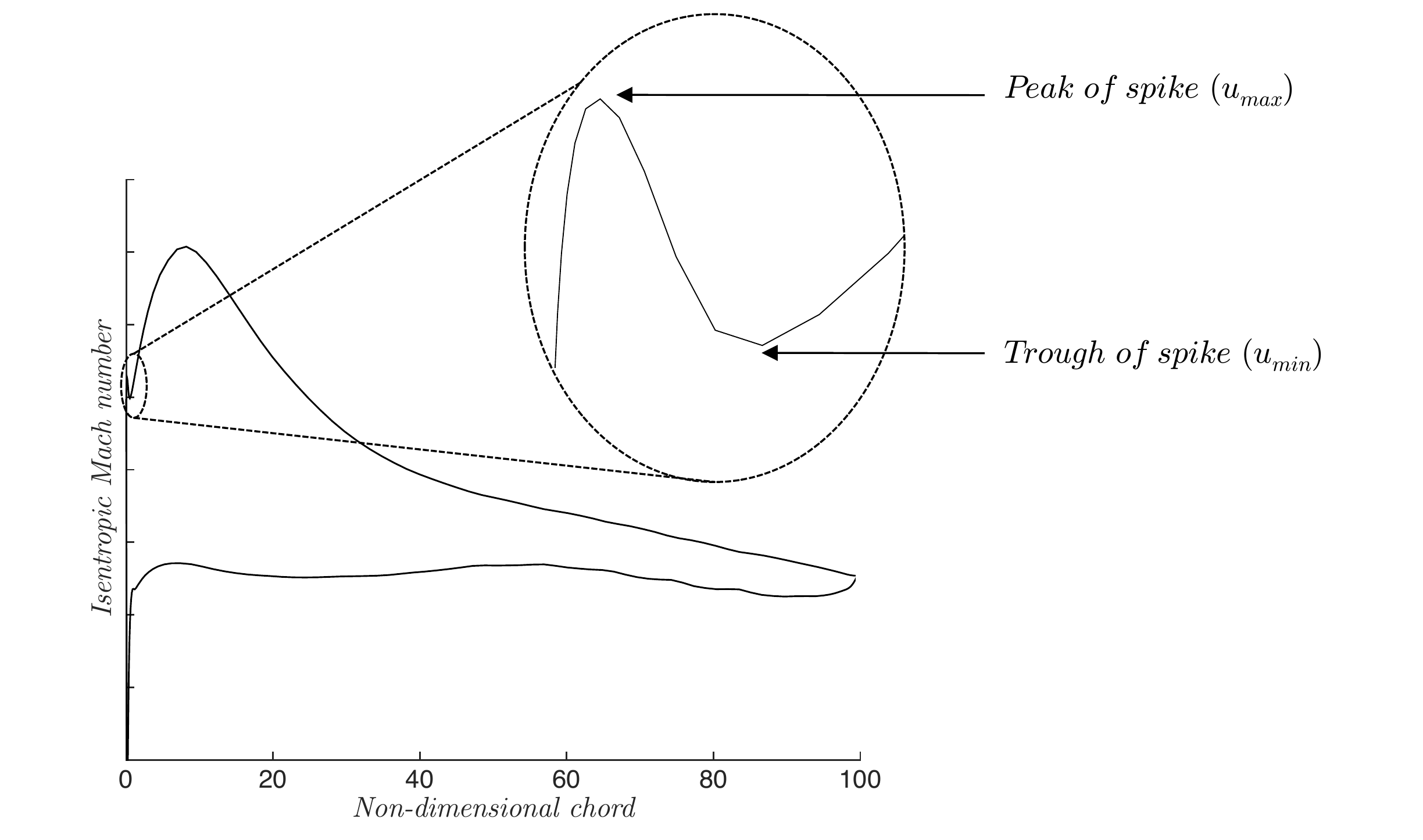}}
  \caption{Schematic showing the suction-side compressor spike on an airfoil}
\label{fig:spikes}
\end{figure}

Compressor spikes \cite{Goodhand_spikes} are a measure of the diffusion that occurs at the leading edge of a fan or compressor blade. The greater the diffusion, the more prone the boundary layers are to separation, causing the flow to transition in the shear layer, thereby increasing profile loss \cite{Goodhand_thesis}. These spikes are characterized by the leading edge spike height, $D_{spike} = (u_{max} - u_{min})/u_{min}$, where $u_{min}$ and $u_{max}$ are the peak and trough velocities of the leading edge spike (defined as per figure \ref{fig:spikes}). The larger the value of $D_{spike}$, the greater the profile loss.
The spike height is determined by the geometry of the compressor leading edge, the Mach number, Reynolds number and the blade loading. The latter affects the spike height through a change in the local incidence, which may be achieved either by changes in the airfoil camberline distribution, 3D blade parameters or changes along the endwalls. 

\begin{figure}
  \centerline{\includegraphics[natwidth=802, natheight=354, width=11cm]{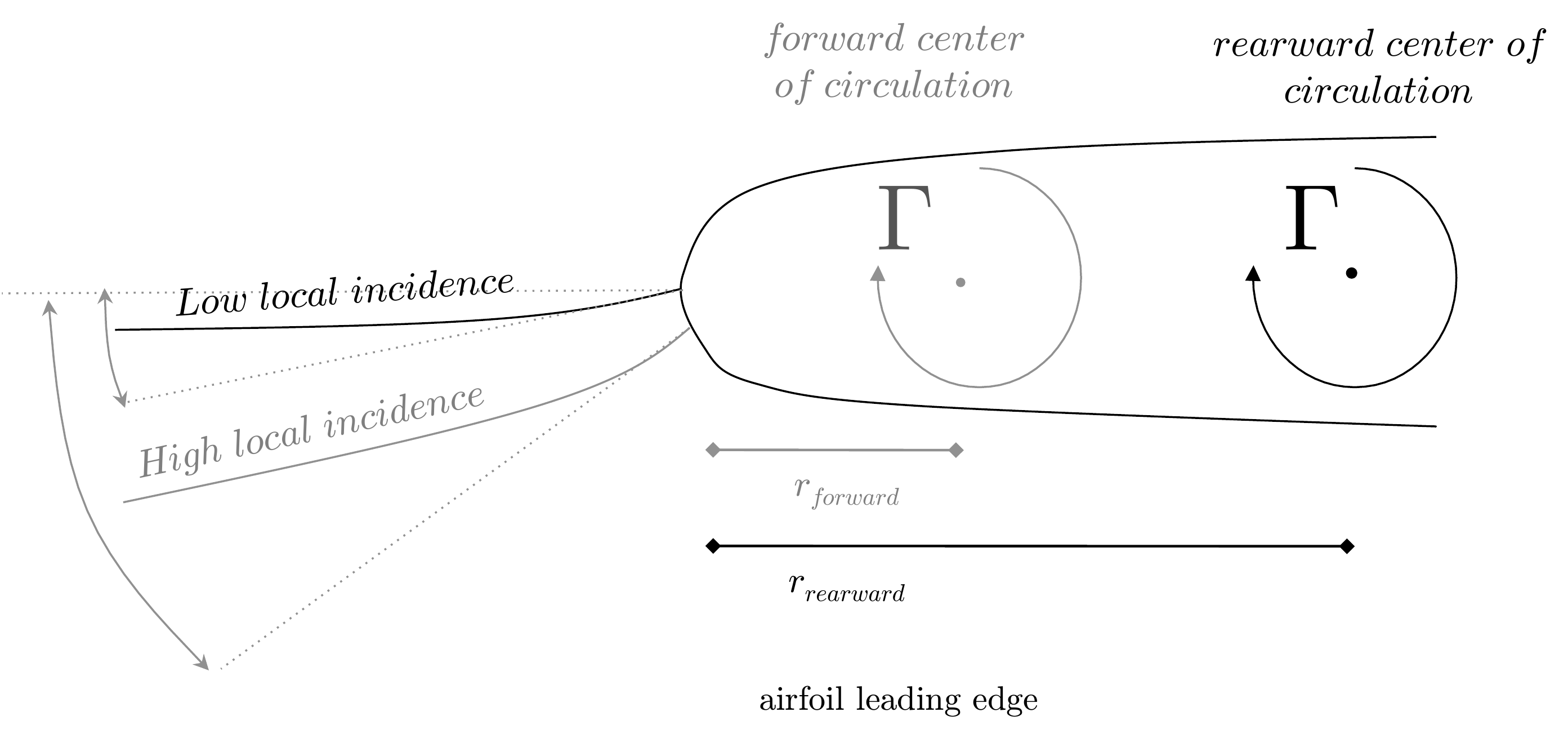}}
  \caption{Schematic showing how a change in the loading distribution affects local incidence, adapted from \cite{Goodhand_thesis}}
\label{fig:leading_edge}
\end{figure}

The effect of a change in loading can be understood as follows. Assume that the loading distribution along the chord is approximated by a single vortex with circulation, $\Gamma$, close to the leading edge, as shown in figure \ref{fig:leading_edge}. The resulting upwash velocity at the leading edge can then be given by
\begin{equation}
u_{LE}=u_{\infty}+\frac{\Gamma}{2\pi r}cos\left(\theta\right)
\end{equation}
where $u_{\infty}$ is the freestream velocity, $r$ is the distance from the center of circulation to the leading edge and $\theta$ is the angle between the camberline at the leading edge point and $r$. The lower the upwash velocity, the lower the incidence. Thus, as the center of loading shifts rearward, $r$ increases causing the upwash velocity to decrease, thereby decreasing the incidence and thus the size of the spike. It should be noted that there is a difference between the behavior of the suction-side spike and the pressure-side one. As the local incidence increases, the suction-surface spike increases but the pressure-surface spike decreases. This is because with increasing incidence the stagnation point is displaced further away from the leading edge, and thus the flow needs to be accelerated hard over the curved airfoil nose to channel the flow through the passage. A similar argument holds if the magnitude of $\Gamma$ closer to the leading edge is larger, as it results in a larger $u_{LE}$ and thus greater incidence \cite{Goodhand_thesis}. 

\subsection{3D fluid mechanics}
Current fan stator designs have benefitted significantly from prior aerodynamic research. Of particular importance is the use of 3D blading methods, such as sweep (airfoil movement parallel to the chordline) and dihedral (airfoil movement perpendicular to the chordline), that have been investigated by \cite{Gallimore_1, Gallimore_2}, \cite{Gummer}, \cite{Sasaki}, \cite{Place_thesis} and  \cite{Bolger_thesis}, \cite{Shahrokh_design1, Shahrokh_design2} to name but a few. Analogous to a swept aircraft wing, sweep in a fan stator delays the formation of shock waves in transonic flows. It also alters the streamline at mid passage creating a twisted stream-surface between adjacent blades. This generates a large secondary flow vortex that covers the full blade passage, and acts against the classical cross flow at the hub, but fosters it at the casing. In his work on fan stage aerodynamics, \cite{Zamboni_thesis} found this sweep-induced vortex to play a role in reducing the effect of the overturning boundary layer, and thus loss on the fan stator. 

\begin{figure}
  \centerline{\includegraphics[natwidth=962, natheight=422, width=14cm]{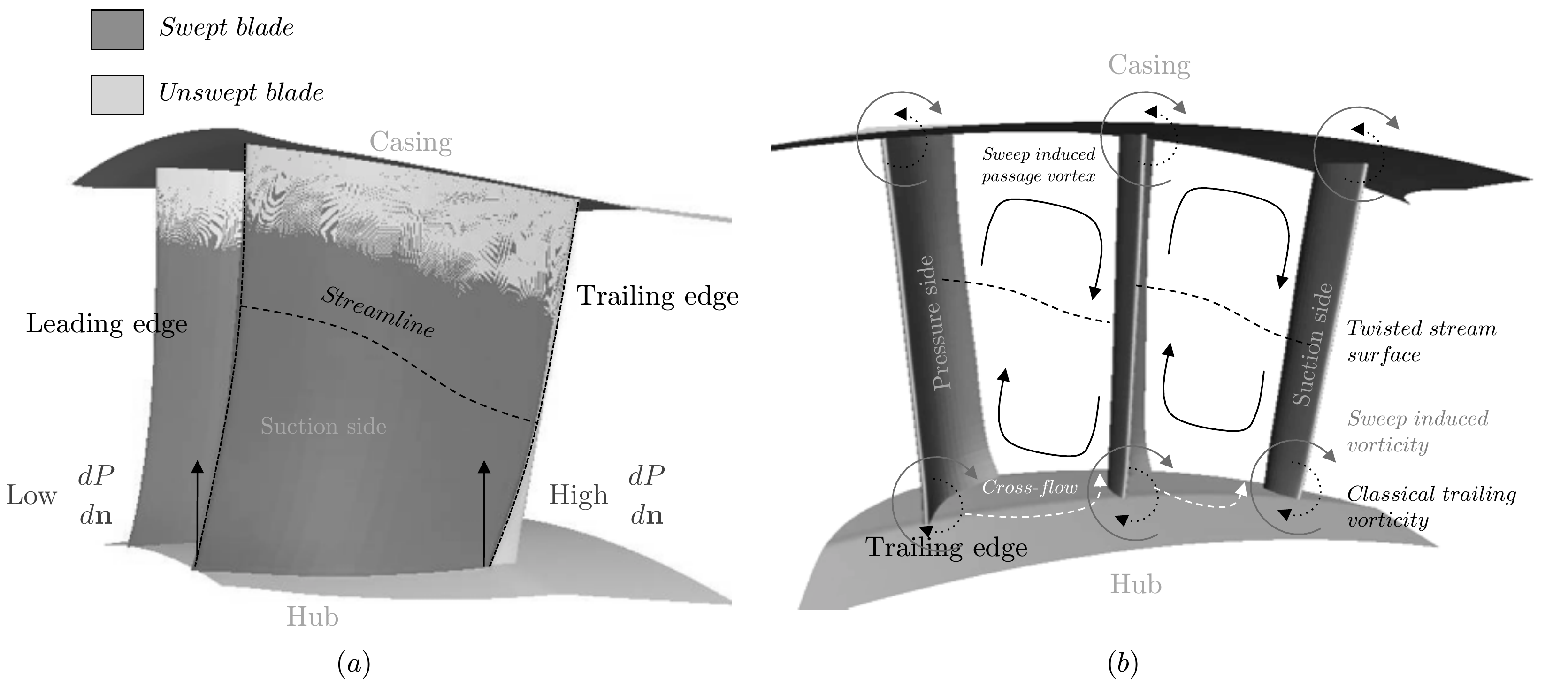}}
  \caption{The effects of sweep in an axial compressor: (a) Change in the hub loading distribution for a swept blade (adapted from \cite{DentonXu}); (b) Change in the 3D stream-surface (adapted from \cite{Gummer})}
\label{fig:sweep}
\end{figure}

Another key effect of sweep is in the reduction of endwall losses. Experiments and numerical results \cite{Gallimore_2} have shown a positive sweep at the hub to reduce secondary flow losses. \cite{DentonXu} attribute this to reduced loading at the leading edge due to the lack of an airfoil section directly above to sustain the pressure rise. This forward movement of the hub airfoil section into a higher pressure region acts to reduce its incidence and peak velocity \cite{Gallimore_2}. The converse is true for the hub trailing edge, where the loading increases due to the stronger pressure gradient enforced by the highly loaded mid-span region (see figure \ref{fig:sweep} for illustrations of both these effects). 

Studies have also shown that a positive dihedral at the hub increases the axial velocity near the trailing edge, reducing loss. The dihedral degree of freedom is analogous to leaning the lifting line, which generates induced velocities as determined by the Biot-Savart law \cite{Gallimore_2, Anderson}. Thus, while positive dihedral is favorable, it does increase the loading on the mid-span sections, which may require a change in the mid-span blade profile. 

\subsection{Effect of leakage flows}
To the best of the authors' knowledge, the only paper that investigates the effect of fan rear-seal leakage flows is that by \cite{Zamboni} where the authors computationally analyze the effect of fan rear-seal leakage with experiments obtained from a high bypass ratio rig. This work is well-motivated, as over time leakage paths do tend to widen in an engine while the leakage flow rate and stagnation temperature tend to rise. This leads to increased secondary flow losses and increased leakage flow temperatures requiring the compressor to do more work to compress the air for the same pressure ratio. In the paper, the authors found rear-seal flow to increase the blockage to the fan stator and lead to a reduction in the capacity ($\Gamma=\dot{m}\sqrt{T_{03}}/P_{03}$), thereby affecting the thrust. Here $\dot{m}$ is the stator mass-flow, and $T_{03}$ and $P_{03}$ are the total temperature and pressure at the stator inlet. They also found that the leakage flow loss can be reduced with a reduced hub annulus line curvature. 

\begin{figure}
  \centerline{\includegraphics[natwidth=560, natheight=420, width=9cm]{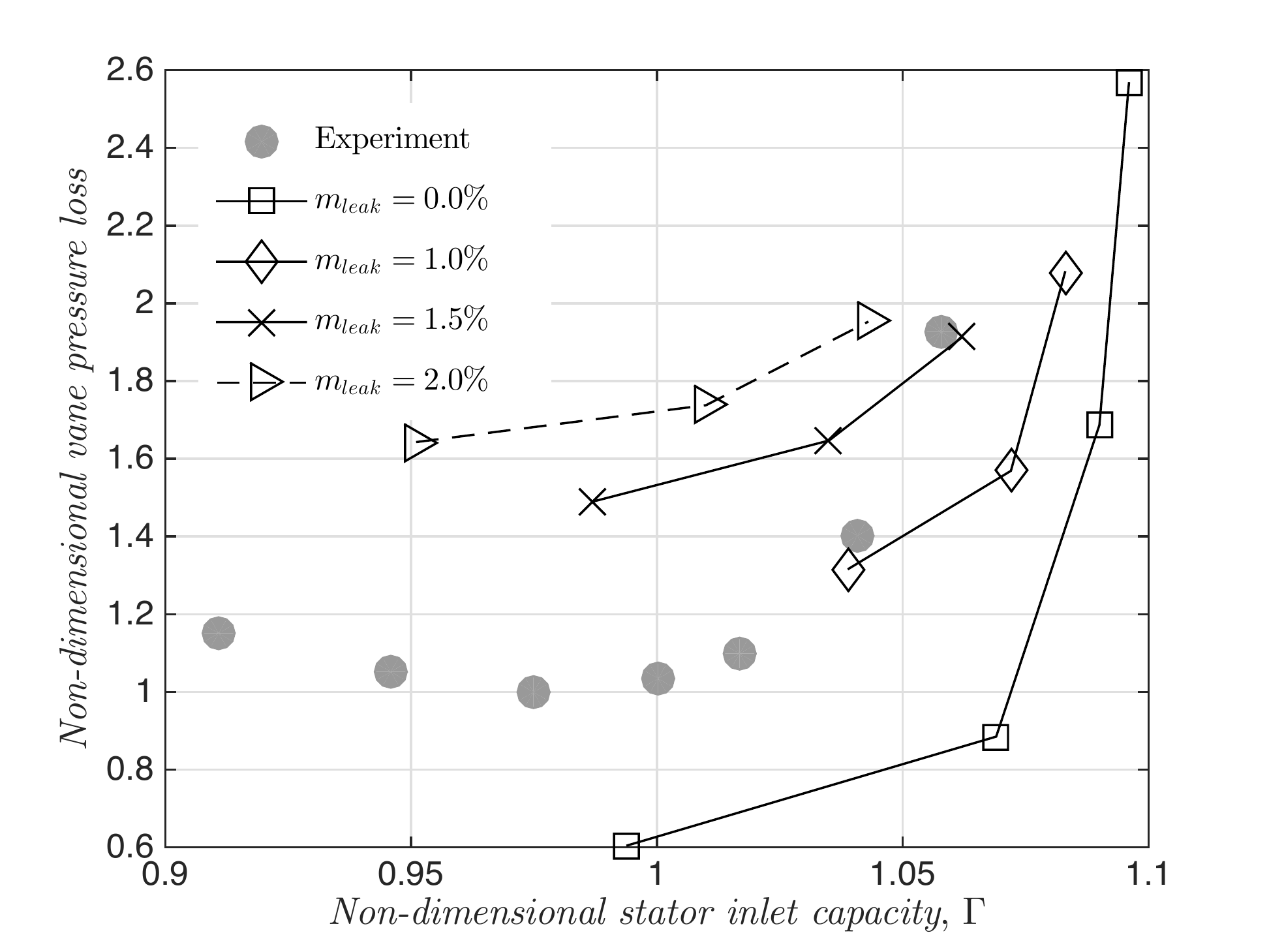}}
  \caption{Simulations seeking to account for the effect of uncertain leakage flow in high bypass fan rig experiments, adapted from \cite{Zamboni}}
\label{fig:zamboni}
\end{figure}

The importance of uncertainty quantification in investigating leakage flows can be understood by analyzing figure \ref{fig:zamboni}, adapted from \cite{Zamboni}. This graph plots experimental and computational values of the loss-loop for a high bypass ratio fan rig. Here, the $y$-axis is the stator pressure loss, which is defined as $\left(P_{03}-P_{05}\right)/\left(P_{03}-P_{3}\right)$, while the $x$-axis represents the flow capacity through the stator. Here and elsewhere in this paper, $P_{3}$ is the static pressure at the stator inlet, while $P_{05}$ is the stagnation pressure at the stator outlet. The numerical predictions, shown as markers joined by lines, were obtained by running the CFD at different leakage mass-flow rates. It is likely that the uncertainty in the experimental leakage flow rate (which cannot easily be measured) is responsible for the scatter when compared with the numerical characteristic lines. It is precisely this result that motivated the robust design optimization study carried out by the authors. 

\section{Computational tools}
Here a brief discussion on the flow solvers used and optimization strategy adopted is provided.

\subsection{Flow simulations}
The aforementioned computational optimization study was carried out using Rolls-Royce's 3D RANS solver HYDRA \cite{Lapworth} on a multi-block structured mesh from PADRAM \cite{Shahrokh2, Shahrokh}. The PADRAM-HYDRA suite has been previously validated by the authors for transonic compressor flows in \cite{Seshadri_LEAK}. Nonlinear flow solves were carried out using HYDRA. HYDRA is an unstructured solver that uses an edge-based data structure with the flow data stored at the cell vertices. A flux-differencing algorithm based on monotone upstream-centered schemes for conservation laws (MUSCL) \cite{vanLeer} is used for space discretization. For the steady-state solution, the pre-conditioning of the discrete flow equations is applied using a block Jacobi pre-conditioner and a five-stage Runge-Kutta scheme. An element-collapsing multi-grid algorithm is used to accelerate the convergence to steady state. In the present study, the Spalart-Allmaras turbulence turbulence model is used, assuming fully turbulent boundary layers (i.e.~no transition modeling) and with wall functions. 

A few points on the boundary conditions employed are in order. Both the stator exit and the bypass used a subsonic exit capacity boundary condition. The interface between the rotating and stationary zones was modeled with a mixing plane boundary condition that computes mass-average values at the rotor-core exit and feeds them to the stator inlet boundary condition. The fan rotor cavity was prescribed with a subsonic inlet whirl massflow boundary condition. Flow solves carried out on the geometry revealed $y^{+}$ values close to 1. Convergence was achieved after approximately 900 iterations. 

Adjoint flow solves were also carried out using HYDRA with Spalart-Allmaras and with wall functions. The adjoint code is based on the discrete adjoint, which involves linearizing the set of discrete Navier-Stokes equations. This linearization can be performed either manually or via a suitable automatic differentiation package \cite{Giles}. Each matrix in the resulting set of equations is then transposed and all the operations are performed in reverse \cite{Mavriplis}. The main advantage of the discrete approach over the continuous one is that it aids in more straightforward code development \cite{Giles}. Another advantage of the discrete adjoint is that when solved exactly, the resulting solution yields an exact gradient when compared with finite differences on the same mesh \cite{Nadarajah}. Typically the adjoint code requires 20-30$\%$ more memory than the non-linear code and costs 10-20$\%$ more time \cite{Giles}. 

\begin{figure}
  \centerline{\includegraphics[natwidth=735, natheight=600, width=14cm]{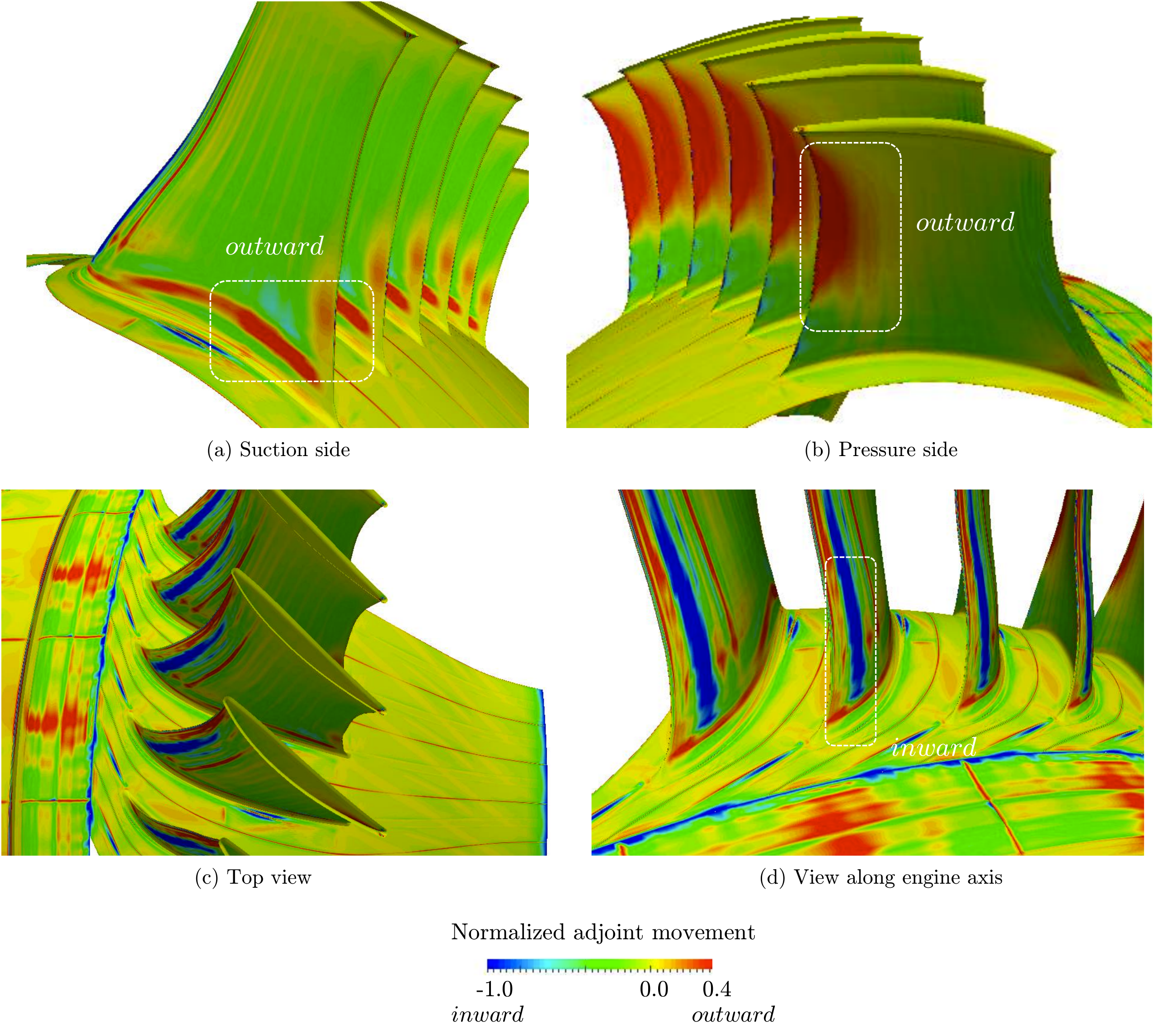}}
  \caption{Contours of the adjoint-efficiency sensitivity}
\label{fig:adj}
\end{figure}

In HYDRA the adjoint equations are solved with a time-marching method that employs partial updates of the numerical smoothness and viscous fluxes with Jacobi preconditioning at select stages in the Runge-Kutta iteration. The scheme ensures exact equivalence with a linear perturbation code during convergence \cite{Giles}. A quick preview of the adjoint flow solution with respect to a certain functional will show the wake starting from the leading edge traveling upstream to the inlet. This \textit{backward} adjoint flow solution is then used to compute the surface sensitivities with each mesh node. Surface sensitivities for the adjoint-efficiency functional for the fan system are shown in figure \ref{fig:adj}. The adjoint surface map motivated the choice of the design parameterization. 
%
%

\subsection{Design space}
\begin{figure}
\centering
\includegraphics[natwidth=962,natheight=362,width=13cm]{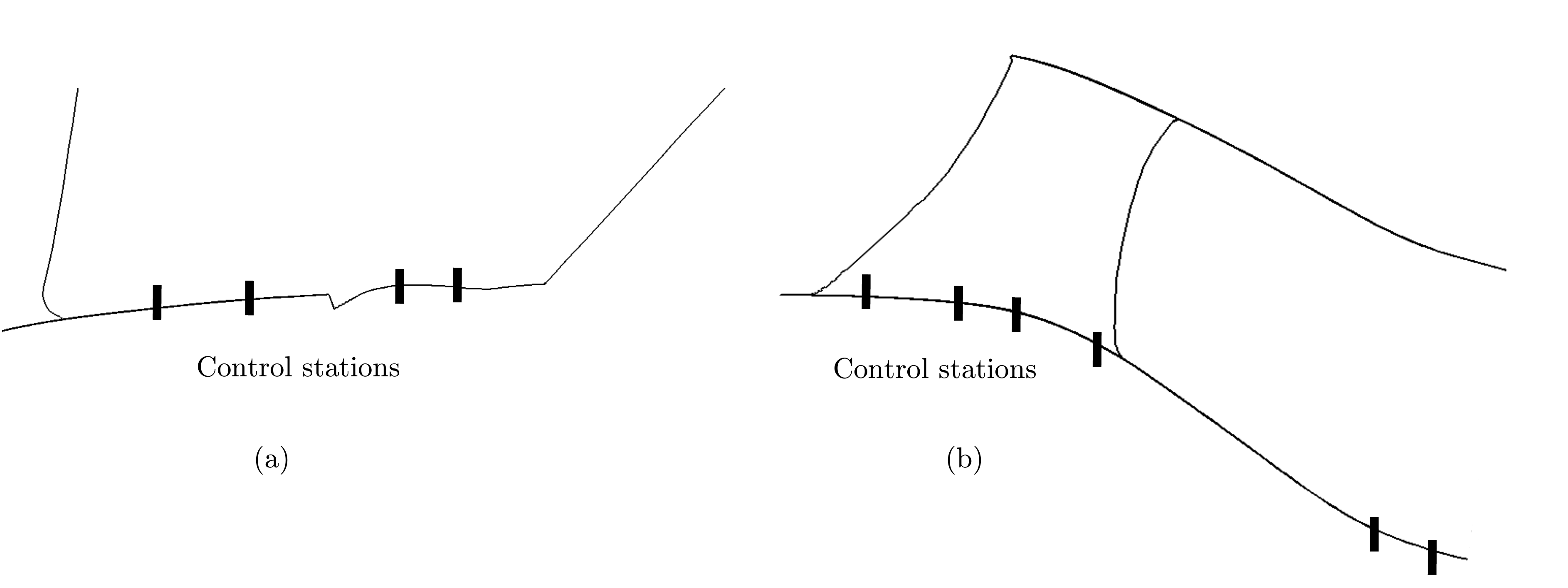}
\caption{Control point locations used for endwall perturbation via univariate b-spline curves: (a) For the rotor hub; (b) For the stator hub.}
\label{fandesign}
\end{figure}

A total of 35 design parameters were used to parameterize the fan stage geometry. Specifically, the focus of interest was the fan rotor downstream hub line, the fan stator hub line and the stator 3D design. The fan rotor geometry was not altered. The stator 3D parameterization is based on 3D blade engineering parameters. This includes tangential lean, skew, leading and trailing edge re-cambering and axial sweep (see \cite{Gallimore_1} for definitions of these degrees of freedom). The aforementioned five degrees of freedom are prescribed at 5 radial stations corresponding to 0,25,50,75,100 percent span, and then linearly interpolated to create the resulting blade profile. This yielded a total of 25 design variables for the 3D stator design. The endwall design perturbations used were based on a radial displacement of a chosen control point, that is then smoothly merged with the undisplaced points on the endwall geometry via a series of univariate b-spline curves. A total of four control stations were chosen immediately aft of the fan root, adjacent to the cavity, and a total of six control stations were chosen along the stator endwall (see Figure~\ref{fandesign}). For the latter, four stations were placed in-line with the stator, and two points were placed downstream. It should be noted that all six perturbations were axisymmetric. 

Using the adjoint method, the sensitivities of any functional with respect to the design variables can be obtained. For the fan system, 35 perturbed meshes, corresponding to the perturbation of each design parameter, were generated. Each mesh was perturbed by adding an $\epsilon$ value to the value of the respective design parameter. The geometric changes in the perturbed mesh were then converted to a finite difference approximation between the two volume meshes (the baseline and the perturbed). The corresponding change in the functional was computed using the adjoint dot product. A key step in this process is the determination of an appropriate $\epsilon$ value for each design parameter. To ensure that the chosen $\epsilon$ value was adequate a few numerical tests were carried out. The efficiency sensitivity with respect to the design parameters for different $\epsilon$ values was computed. Different sets of $\epsilon$ values were used for the three design parameterizations. Both the fan and stator hub geometries were found to yield comparable gradients for $\epsilon$ values in the range of 0.0003 to 0.001, while for the stator 3D parameters this range was 0.01 to 0.1. Values below this range typically yielded gradients with a lot of noise, and values higher were neglected because the step size was too large relative to the design parameter range. 

\subsection{Optimization strategy}
To de-sensitize the aerodynamics to the effect of the increasing leakage flow over life, the optimization strategy sought to increase the mean efficiency while decreasing its variance (see Figure~\ref{fig:engine2}). Mean and variance values in the stage efficiency were obtained by running flow both nonlinear and adjoint solves at two rear-seal leakage flow conditions ($\dot{m}_{leak,low} = 0.13\%$ and $\dot{m}_{leak, high} = 0.30\%$) and assuming a certain probability distribution in the rear-seal leakage flow to characterize its uncertainty. After running the flow solvers, the corresponding probability distribution in the stage efficiency could be computed. For both rear-seal leakage flow conditions, the stage efficiency was computed by
\begin{equation}
\eta_{root}=\left(\frac{PR^{\left(\gamma-1\right)/\gamma}-1}{TR-1}\right),
\end{equation}
with,
\begin{equation}
PR=\left(\frac{P_{05}}{P_{01}}\right) , \; \; TR=\left(\frac{T_{05}}{T_{01}}\right),
\end{equation}
where the individual pressures and temperatures are mass-averaged circumferentially and then radially. Note that subscript $5$ denotes the stator outlet; as mentioned earlier, and subscript $1$ denotes the fan inlet. Here $\gamma$ is the universal gas constant. Specifics on the optimization method, the objective and gradient formulations are beyond the scope of this paper and can be found in \cite{Seshadri_ASME2}. The focus of this paper is solely to understand why the optimized design performed better as it achieved both a higher mean efficiency and a decrease in the variance. 

\section{Robust design aerodynamics}
This section discusses the aerodynamics of the optimized design with respect to the initial design. As mentioned earlier, the optimized design was achieved using the technique in \cite{Seshadri_ASME2}, and is not elaborated upon here. It should, however, be noted that the technique makes use of both nonlinear and adjoint flow solves at the low- and high-leakage conditions. 

\subsection{Results and analysis}
\begin{figure}
  \centerline{\includegraphics[natwidth=663, natheight=445, width=13cm]{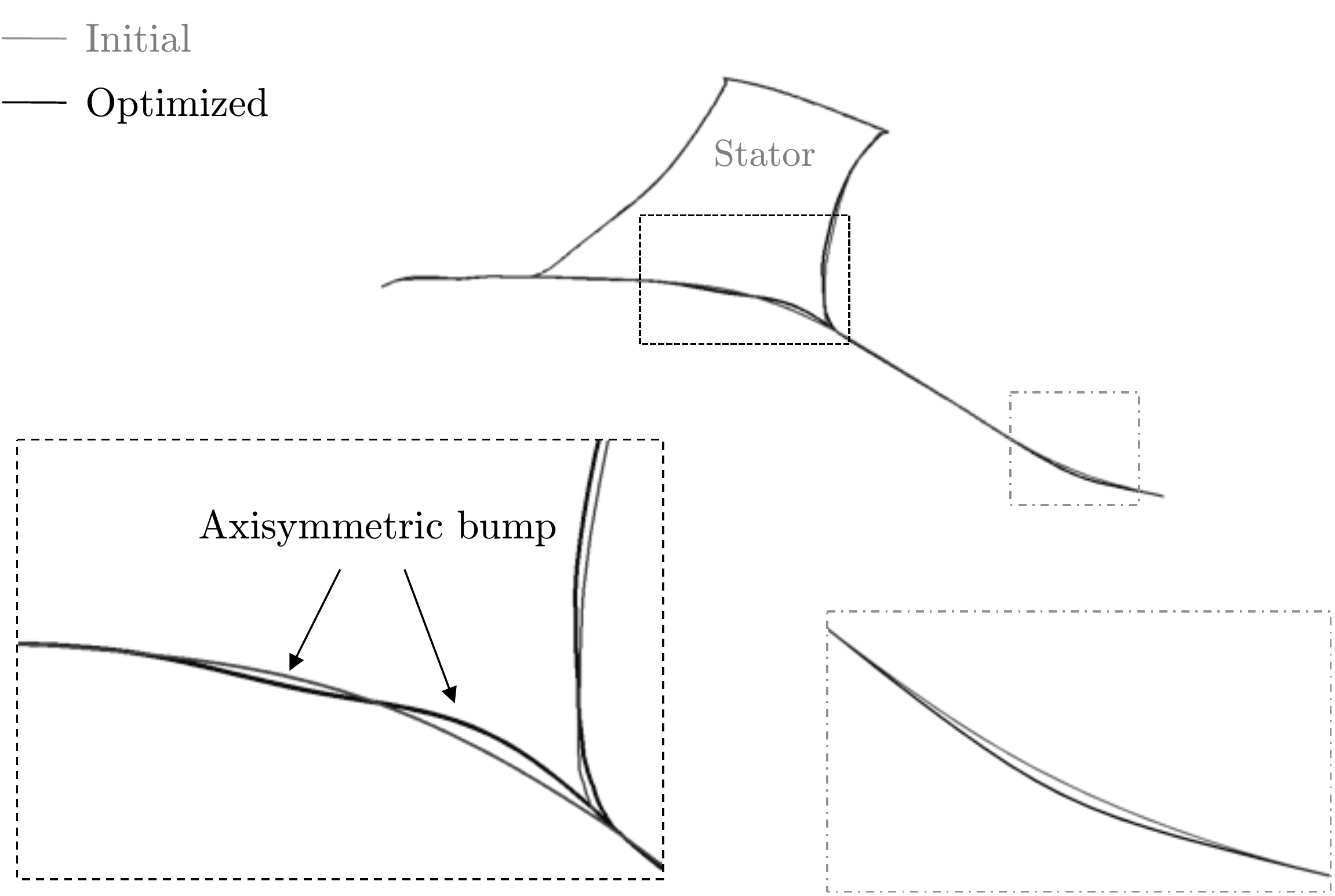}}
  \caption{Meridional view of the profiles of the initial and optimized geometries}
\label{fig:merid}
\end{figure}

\begin{figure}
  \centerline{\includegraphics[natwidth=844, natheight=300, width=13cm]{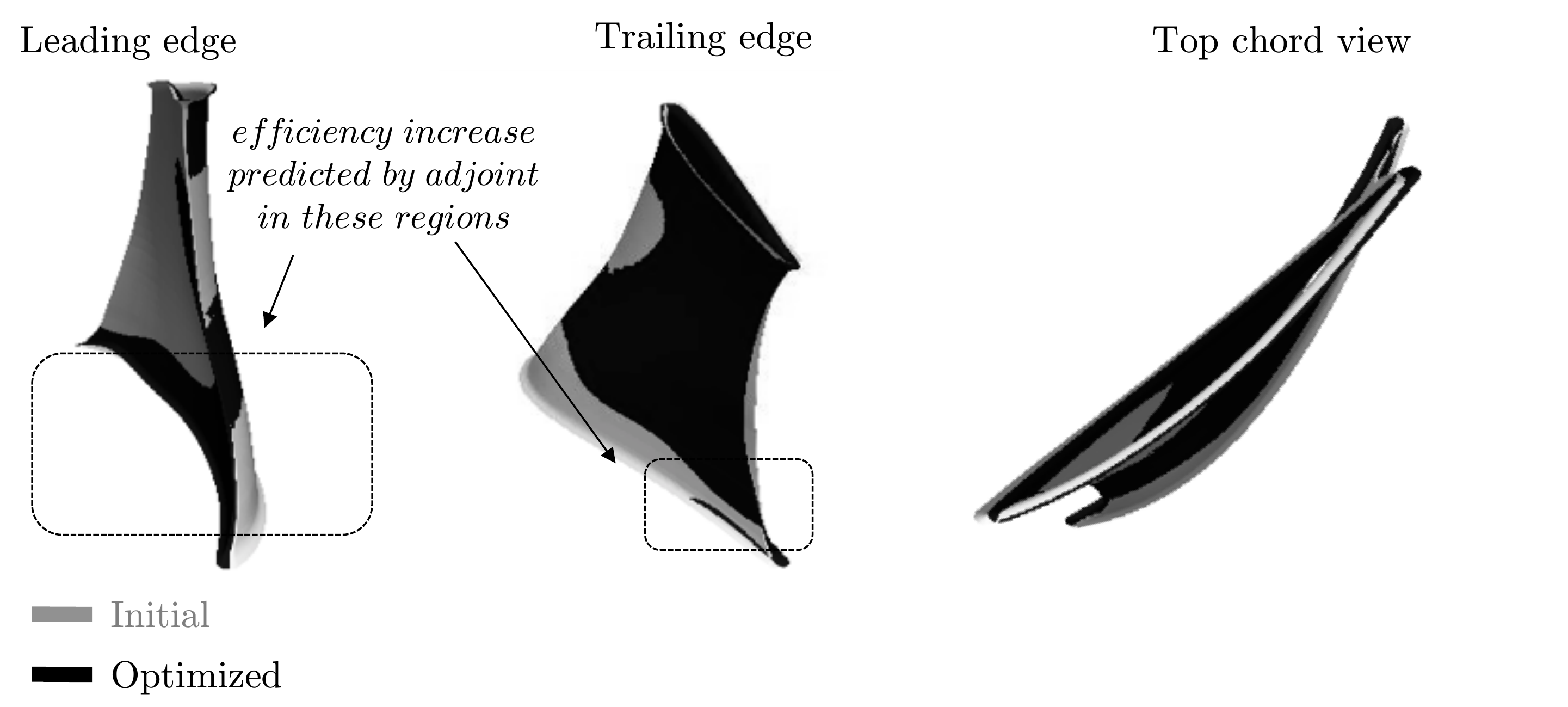}}
  \caption{Stator 3D views of the initial and optimized geometries}
\label{fig:stator3d}
\end{figure}

A comparison of the initial and optimized designs from figure \ref{fig:old_result} are shown in figures \ref{fig:merid} and \ref{fig:stator3d}. Here figure \ref{fig:merid} plots meridional views of the geometries, while figure \ref{fig:stator3d} plots 3D snapshots of the stators. Compared to the initial design, the changes in the optimized design can be summarized as follows:
\begin{enumerate}
\item there is a visible depression along the stator hub line at 60$\%$ chord and an elevation in the hub line at $80\%$ chord;
\item the stator hub section has positive leading edge re-cambering and positive sweep;
\item the stator hub towards the trailing edge has been visibly extruded. 
\end{enumerate}
It is interesting to note that (ii) and (iii) are clearly captured by the adjoint sensitivity maps. Thus, in addition to playing a role in de-sensitizing the stage efficiency to leakage flow, which will be analyzed in subsequent paragraphs, they have an effect on increasing the mean efficiency. 

\begin{figure}
  \centerline{\includegraphics[natwidth=830, natheight=374, width=14cm]{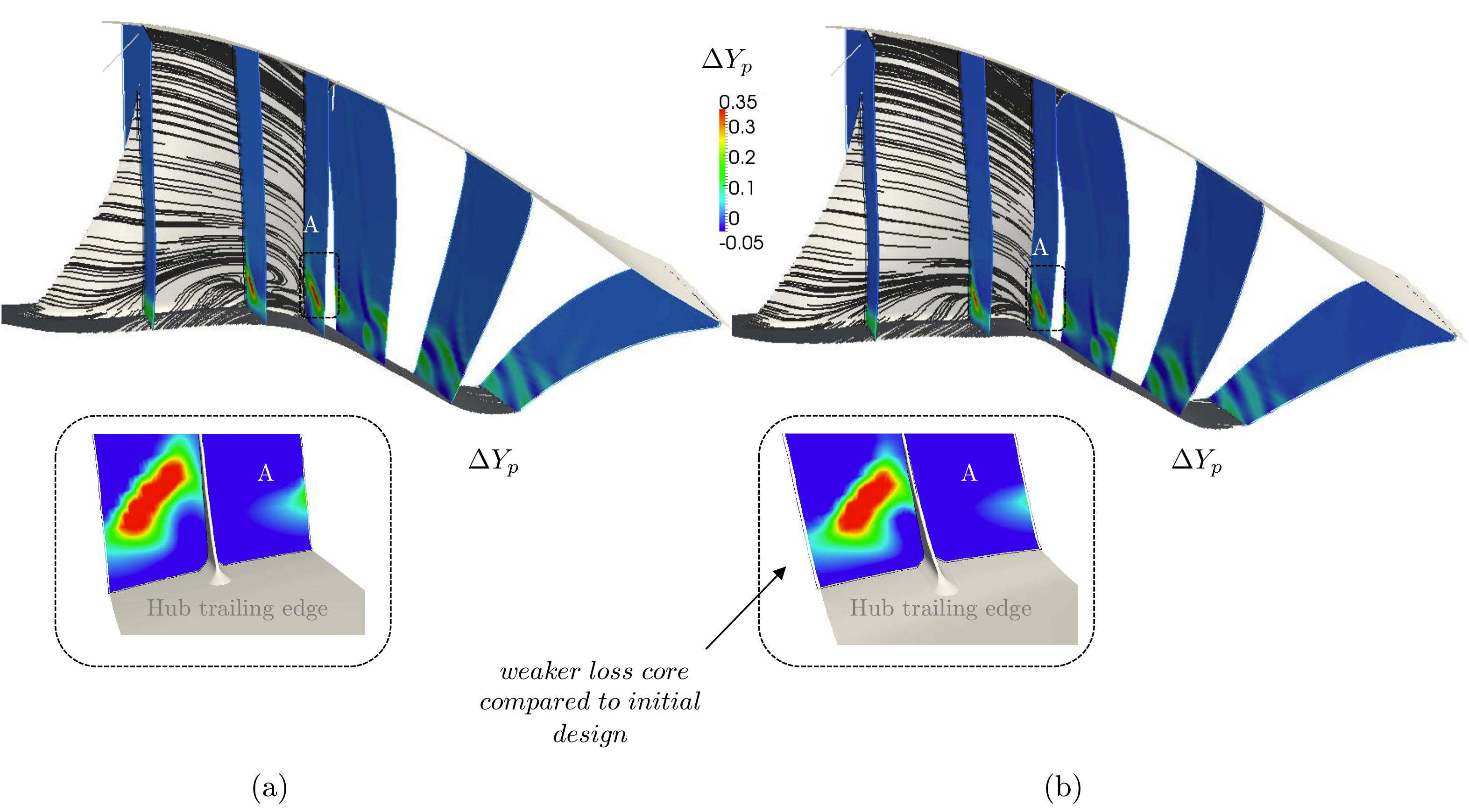}}
  \caption{Surface streamlines with contours of $\Delta Y_p$ for the initial design in (\textit{a}) and optimized design in (\textit{b})}
\label{fig:flowfields}
\end{figure}

The effect of increased sweep at the hub can be understood from the schematic shown in figure \ref{fig:sweep}. As mentioned earlier, an increase in the sweep results in a sweep-induced passage vorticity, which opposes the vorticity associated with the overturned boundary layer in the passage. To understand the effect of the other geometry changes, figure \ref{fig:flowfields} plots the streamlines with slices of the delta pressure loss coefficients on the two geometries. The delta pressure loss coefficient, is defined as
\begin{equation}
\Delta Y_{p}=Y_{p_{\dot{m}_{leak} \; \; at \; \; 0.13} }-Y_{p_{\dot{m}_{leak} \; \; at \; \; 0.30}},
\end{equation}
where
\begin{equation}
Y_{p_{\dot{m}_{leak} \; \; at \; \; i}}=\frac{P_{03}-P_{0}}{P_{03}-P_{3}}
\end{equation}
for leakage condition $i$. Here $P_{03}$ and $P_{3}$ are the stagnation and static pressures at the stator inlet. 
The surface streamlines in the figure are shown for the high-leakage condition, that is $\dot{m}_{leak} = 0.30\%$. It is clear that the initial design shows signs of minor corner separation, which is caused by the presence of a strong overturning of the boundary layer, attributed to the static pressure difference at the hub between the pressure side of the vane and the suction side of the neighboring vane. By decreasing this cross-passage pressure gradient, the effect of the overturning boundary layer can be reduced. In figure \ref{fig:flowfields}(b), the corner separation is virtually eliminated. Plots of $\Delta Y_p$ also show the relative reduction in loss between the high- and low-leakage cases for the optimized design. An inspection of the $\Delta Y_p$ contours fore of the leading edge of the rotor (not shown here), immediately aft of the leakage cavity, showed negligible differences between the initial and optimized designs. This indicates that the role played by the four hub endwall design parameters on both sides of the cavity had little impact on the optimized design.  

\begin{figure}
  \centerline{\includegraphics[natwidth=830, natheight=374, width=13cm]{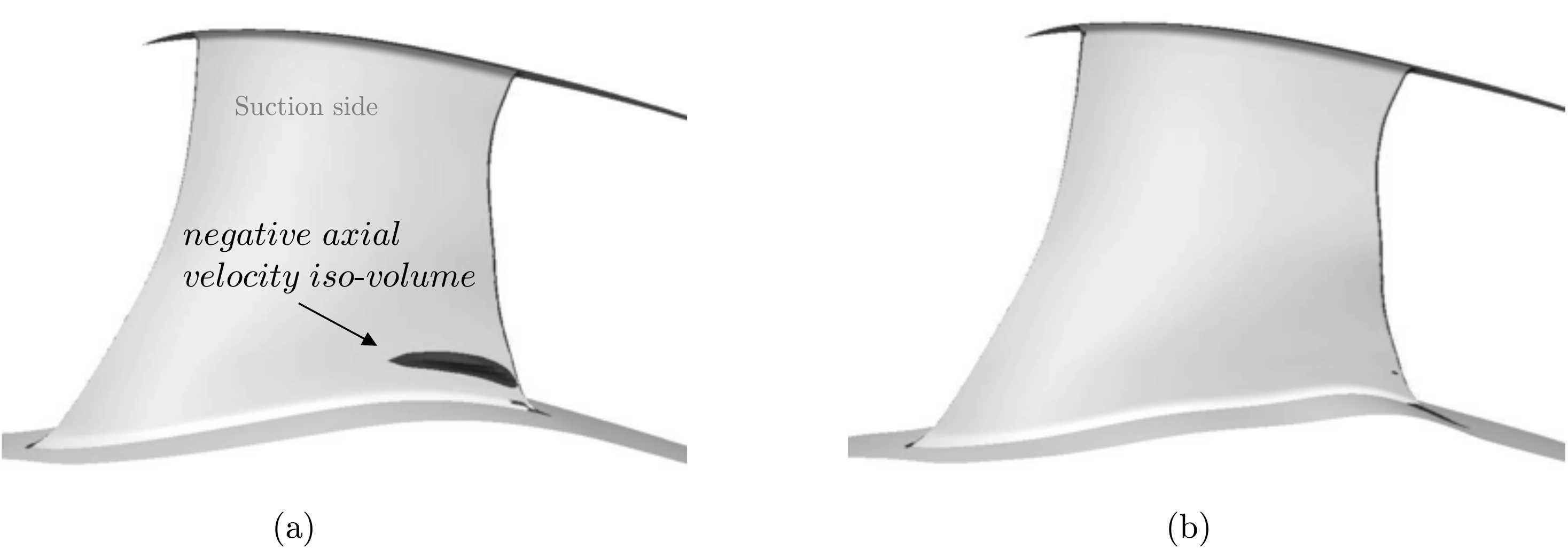}}
  \caption{Iso-volumes of negative axial velocity for the initial design in (\textit{a}) and optimized design in (\textit{b}). These iso-volumes were captured at the high-leakage mass-flow rate condition of 0.30$\%$}
\label{fig:flowfields2}
\end{figure}

The lack of any negative axial velocity on the vane for the optimized design (see figure \ref{fig:flowfields2}) is one of the most significant aerodynamic contributions of this study. Design solutions that can completely mitigate the corner separation, while still maintaining a high aerodynamic efficiency, are of considerable interest. 

\begin{figure}
  \centerline{\includegraphics[natwidth=994, natheight=914, width=13cm]{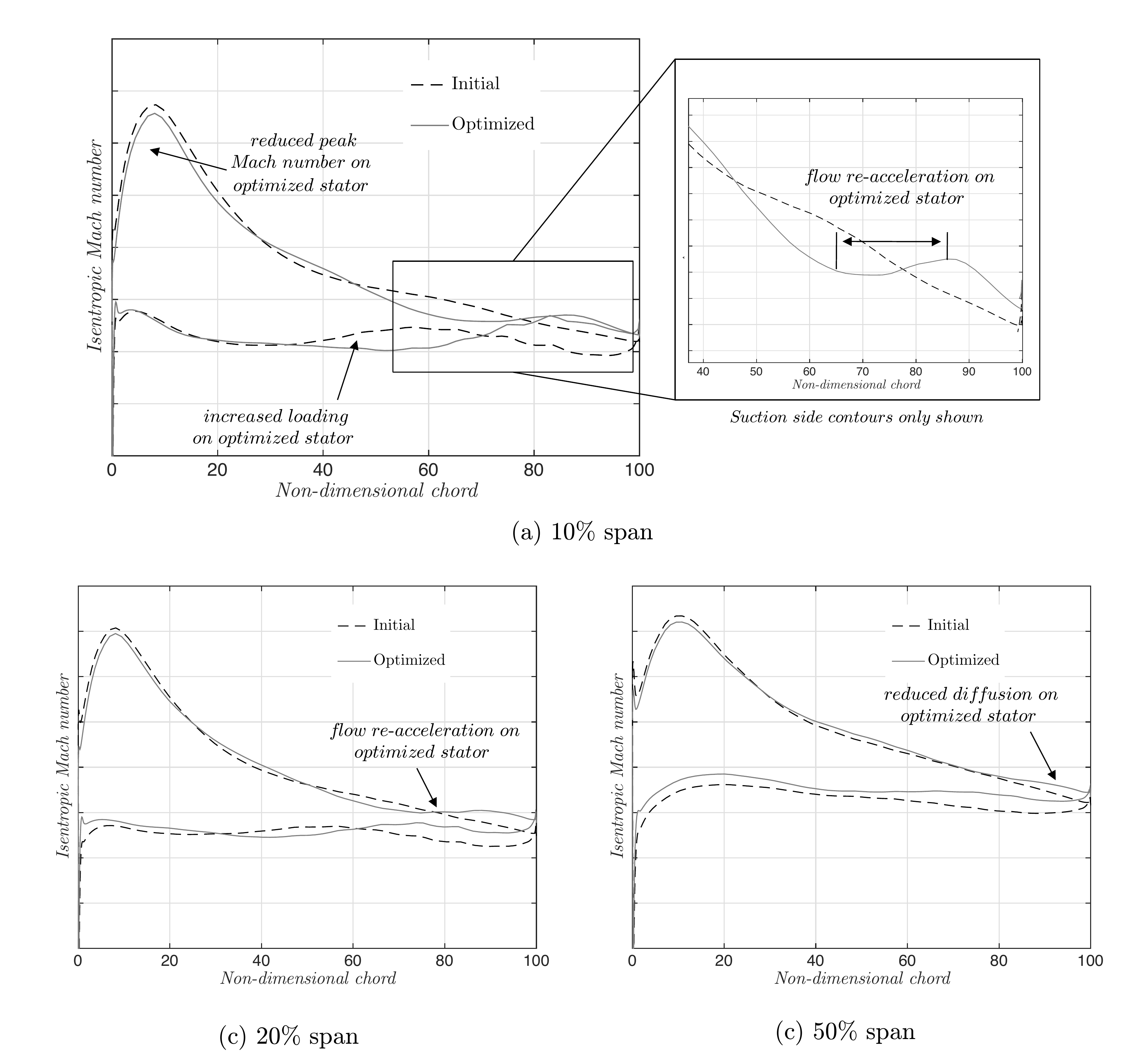}}
  \caption{Isentropic Mach number contours at various spanwise locations for the initial and optimized designs}
\label{fig:isen}
\end{figure}

For a deeper understanding, isentropic Mach numbers of the two geometries are compared in figure \ref{fig:isen}. There are three main observations:
\begin{enumerate}
\item there is a reduction in the suction-side compressor spike for the optimized geometry, while there is an increase in the pressure-side spike throughout the span (see table \ref{table:spike});
\item the optimized geometry exhibits increased loading from $30\%$ to $60\%$ chord, from the hub to $40\%$ span;  
\item the optimized geometry exhibits flow acceleration on the suction side from the hub to $20\%$ span between $65\%$ to $85\%$ chord. 
\end{enumerate}

\begin{table}
  \begin{center}
  \begin{tabular}{lcc}
\hline
Location  &  Initial   &   Optimized \\[3pt]
\hline
       Suction side   & 0.013 & 0.003\\
       Pressure side   & 0.018 & 0.034\\
       \hline
  \end{tabular}
  \caption{Values of suction- and pressure-surface compressor spikes}
  \label{table:spike}
  \end{center}
\end{table}

The change in the spike height of the optimized geometry---relative to the initial design---occurs because of a change in the local incidence at the leading edge of the stator. This, in turn, is attributed to a change in the loading distribution \cite{Goodhand_thesis}. In the optimized geometry there is a rearward shift in the loading and associated reduction in the incidence. This reduces the profile loss at the leading edge of the stator, thus potentially delaying transition. 

The optimized geometry at the hub also exhibits greater diffusion. If the pressure gradient observed at the hub between 45$\%$ to $62\%$ chord were to continue to the trailing edge, the flow would most certainly separate prematurely. However, owing to the interaction of the optimized hub line and stator vane, a region of flow acceleration is observed. The acceleration of the flow in this region results in a thinner turbulent boundary layer which is able to negotiate the adverse pressure gradient more favorably. \cite{Harvey} state that ``...this acceleration is analogous to [the flow] at the front of the airfoil, where a thin boundary layer is able to negotiate the strong diffusion after the peak Mach number point''. In \cite{Harvey}, the flow acceleration was achieved throughout the span by adding metal to the airfoil towards the trailing edge in the form of smoothly blended bumps. Here it is achieved only at the hub due to the stator hub line curvature. How exactly this hub re-acceleration is achieved will be discussed in the next paragraph. However, before the discussion on the isentropic Mach number plots is concluded, there is one more salient point to note.  At and above $50\%$ span there is a reduction in the diffusion on the optimized stator compared with the datum (see figure \ref{fig:isen}(c)). This is favorable as it reduces the risk of flow separation at off-design flow conditions. 

\begin{figure}
  \centerline{\includegraphics[natwidth=940, natheight=399, width=13cm]{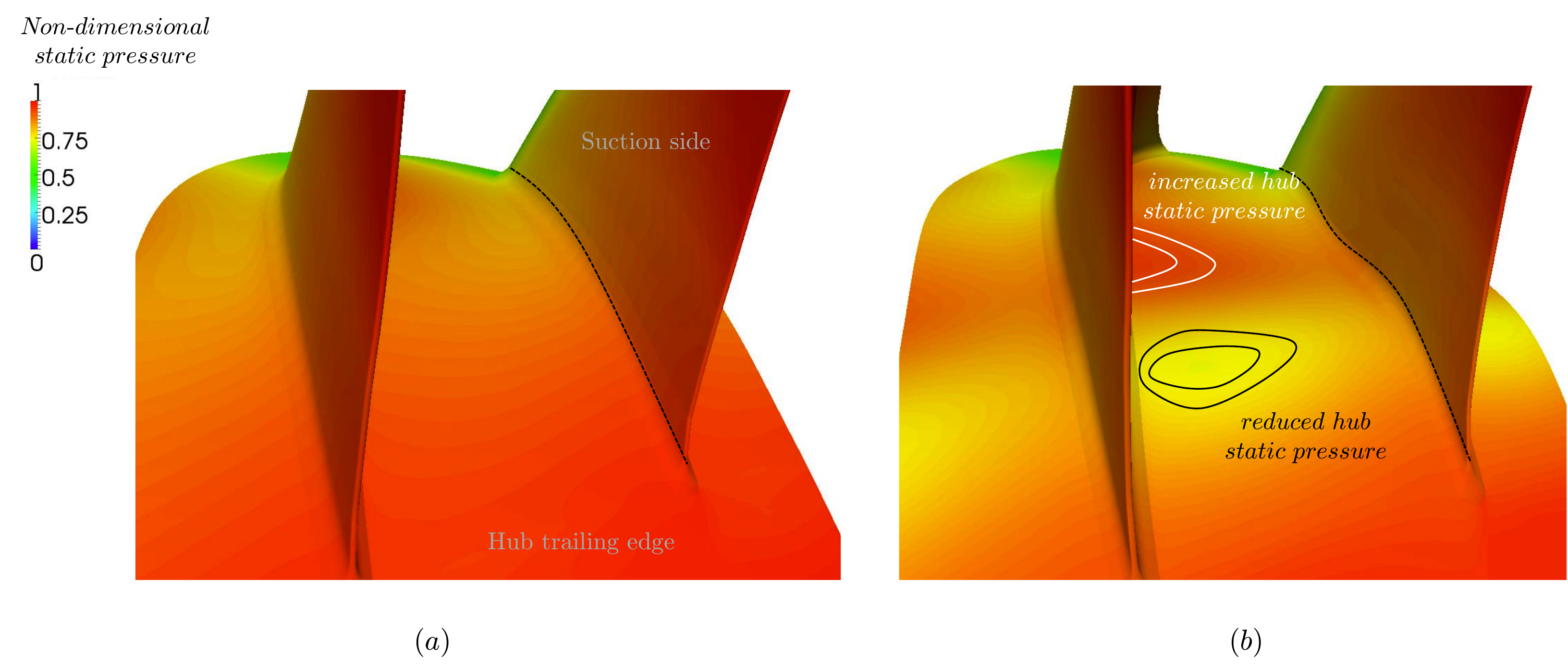}}
  \caption{Comparison of static pressure contours at the hub for the initial design in (\textit{a}) and optimized design in (\textit{b})}
\label{fig:last}
\end{figure}

Figure \ref{fig:last} plots the hub static pressure contours for the initial and optimized designs as seen from the trailing edge. Recall, figure \ref{fig:merid} where the profile of the two geometries was plotted from the meridional perspective. The depression and subsequent elevation observed for the optimized geometry results in a decrease and subsequent increase in the velocity magnitude (mass-flow rate through the passage being constant). The location of the decrease in velocity, and thus increase in hub static pressure, occurs where the cross-passage flow interacts with the suction-side boundary layer. This prevents the cross flow from interacting fully with the suction-side boundary layer. Towards the trailing edge, the elevation of the endwall and subsequent acceleration imparts high-momentum fluid into the corner region, which is typically prone to separation.

\section{Conclusions}
In this paper a detailed aerodynamic analysis of a new fan stage design was undertaken. The design was found to be relatively insensitive to the effect of increasing rear-seal leakage flows. The aerodynamic analysis showed that this insensitivity could be attributed to three main factors: a slight rearward shift in loading, and thus a reduction in incidence; a reduction in the cross-passage pressure gradient; and a re-acceleration of the flow towards the trailing edge, which prevented corner separation. 

\section*{Acknowledgements}
This research was funded through a Dorothy Hodgkin Postgraduate Award, which is jointly sponsored by the Engineering and Physical Sciences Research Council (EPSRC) (UK) and Rolls-Royce plc. The first author would like to acknowledge the financial assistance provided by St.~Edmund's College, Cambridge. The authors would like to thank David Radford and Mark Wilson at Rolls-Royce plc for their assistance in various parts of this work. The authors thank Rolls-Royce plc for permission to publish this manuscript. 

\bibliography{references}
\bibliographystyle{unsrt}
\end{document}